\documentclass[preprint,12pt,3p]{elsarticle}
\usepackage{xspace}
\usepackage{comment}
\usepackage{graphicx}
\usepackage{epstopdf}
\usepackage{epsfig}
\usepackage{subfigure}
\usepackage{url}
\usepackage{amsmath}
\usepackage{listings}
\usepackage{mathtools}
\usepackage{algorithm}
\usepackage[noend]{algpseudocode}
\usepackage{amsfonts}
\usepackage{hyperref}
\usepackage{multirow}
\usepackage{amssymb}
\usepackage{amsmath}
\usepackage{mathrsfs}
\usepackage[utf8]{inputenc}
\newcommand{\XOR}{$\mathbin{\oplus}$}

\journal{Memristive Devices for Brain-inspired Computing as a book chapter}
\let\today\relax

\begin{document}
\title{Hyperdimensional Computing Nanosystem}

\author[ethz,ucb]{Abbas Rahimi}
\ead{abbas@ee.ethz.ch}
\author[stanford]{Tony F. Wu}
\ead{tonyfwu@stanford.edu}
\author[stanford]{Haitong Li}
\ead{haitongl@stanford.edu}
\author[ucb]{Jan M. Rabaey}
\ead{jan@eecs.berkeley.edu}
\author[stanford]{H.-S. Philip Wong}
\ead{hwangw@stanford.edu}
\author[mit]{Max M. Shulaker}
\ead{shulaker@mit.edu}
\author[stanford]{Subhasish Mitra}
\ead{subh@stanford.edu}

\address[ethz]{ETH Zurich}
\address[ucb]{UC Berkeley}
\address[mit]{MIT}
\address[stanford]{Stanford}

\begin{abstract}
One viable solution for continuous reduction in energy-per-operation is to rethink functionality to cope with uncertainty by adopting computational approaches that are inherently robust to uncertainty.
It requires a novel look at data representations, associated operations, and circuits, and at materials and substrates that enable them.
3D integrated nanotechnologies combined with novel brain-inspired computational paradigms that support fast learning and fault tolerance could lead the way.
Recognizing the very size of the brain's circuits, hyperdimensional (HD) computing can model neural activity patterns with points in a HD space, that is, with \emph{hypervectors} as large randomly generated patterns.
At its very core, HD computing is about manipulating and comparing these patterns inside memory.
Emerging nanotechnologies such as carbon nanotube field effect transistors (CNFETs) and resistive RAM (RRAM), and their monolithic 3D integration offer opportunities for hardware implementations of HD computing through tight integration of logic and memory, energy-efficient computation, and unique device characteristics.
We experimentally demonstrate and characterize an end-to-end HD computing nanosystem built using monolithic 3D integration of CNFETs and RRAM.
With our nanosystem, we experimentally demonstrate classification of 21 languages with measured accuracy of up to 98\% on $>$20,000 sentences (6.4 million characters), training using one text sample ($\approx$100,000 characters) per language, and resilient operation (98\% accuracy) despite 78\% hardware errors in HD representation (outputs stuck at 0 or 1).
By exploiting the unique properties of the underlying nanotechnologies, we show that HD computing, when implemented with monolithic 3D integration, can be up to 420$\times$ more energy-efficient while using 25$\times$ less area compared to traditional silicon CMOS implementations.
\end{abstract}
\maketitle


\section{Introduction}
Over the past six decades, the semiconductor industry has been
immensely successful in providing exponentially increasing
computational power at an ever-reducing cost and energy footprint.
Underlying this staggering evolution is a set of well-defined
abstraction layers: starting from robust switching devices that
support a deterministic Boolean algebra, to a scalable and stored
program architecture that is Turing complete and hence capable of
tackling (almost) any computational challenge. Unfortunately,
this abstraction chain is being challenged as scaling continues to
nanometer dimensions, as well as by exciting new applications that
must support a myriad of new data types. Maintaining the current
deterministic computational model ultimately puts a lower bound on
the energy scaling that can be obtained, set in place by fundamental
physics that governs the operation, variability and reliability of
the underlying nanoscale devices~\cite{RoadMapl_to_LV,EnergyScaBnd,var_survey}.

At the same time, the nature of computing itself is evolving rapidly: for a vast number of emerging applications, cognitive functions such as classification, recognition, and learning are rapidly gaining importance.
For efficient information-extraction, these applications require a fundamental departure from the traditional von Neumann architecture, where data has to be transported to the processing unit and back, creating the infamous memory wall.
One of the most promising options for realizing such non-von Neumann architectures is to exploit beyond silicon materials and substrates that allow dense and 3D integration of memory and logic.
However, such a dense and layered 3D system increases the risk of failures within the chip, and system must be fault-tolerant.
As has been realized for a long time, this also resembles the way brain computes.
Hence, 3D integrated nanotechnologies combined with brain-inspired computational paradigms that support fast learning and fault-tolerance could lead the way~\cite{TCAS17}.

Emerging hyperdimensional (HD) computing~\cite{HD09} is based on the understanding that brains compute with \emph{patterns of neural activity} that are not readily associated with scalar numbers.
In fact, the brain's ability to calculate with numbers is feeble.
However, due to the very size of the brain's circuits, we can model neural activity patterns with points of a HD space, that is, with hypervectors.
In this formalism, information is represented in hypervectors as ultra-wide words.
Such hypervectors can then be mathematically manipulated to not only classify but also to bind, associate, and perform other types of cognitive operations in a straightforward manner.
In addition, these mathematical operations also ensure that the resulting hypervector is unique and thus the learning is one-shot or few-shot meaning that object categories are learned from few examples in a \emph{single pass} over the training data~\cite{EMG-HD,ISCAS18,BioCAS18,BICT17,MONET17}.
Thus HD computing can substantially reduce the number of operations needed by conventional learning algorithms, thereby providing tremendous energy savings.
Implementation of the HD computing in practical hardware needs large arrays of non-volatile memory so that the learning is not ``forgotten.''
Our approach is therefore focused on potential low-voltage, non-volatile Resistive Random Access Memory (RRAM) that can be integrated at high density with logic switches~\cite{IEDM_2016}.
We further explore potential low-voltage approaches to logic transistors such as the carbon nanotube field effect transistors (CNFETs) so that the overall supply voltage requirement and hence the energy dissipation can be lowered~\cite{ISSCC2018,JSSC2018}.

In this book chapter, we present HD computing nanosystem by efficient implementation of HD operations using emerging nanoscalable CNFETs and RRAM, and their monolithic 3D integration.
The rest of this book chapter is organized as follows.
In Section~\ref{sec:HD}, we briefly introduce HD computing and discuss some of its key properties including a well-defined set of arithmetic operations (Section~\ref{sec:arith_ops}), generality and scalability (Section~\ref{sec:universal}), robustness (Section~\ref{sec:robustness}), and embarrassingly parallel operations (Section~\ref{sec:memory-centric}).
In Section~\ref{sec:lang_example}, we describe an application of HD computing in language recognition, and show how its operations can be used to solve various learning and classification tasks.
In Section~\ref{sec:emergingTech}, we present the emerging technology for HD computing and describe how the principal operations can be efficiently implemented in a 3D integrated architecture.
Our experimental results for 3D architecture regarding robustness and energy efficiency are described in Section~\ref{sec:expdemo}.

\section{Background in HD Computing}
\label{sec:HD}
The difference between traditional computing and HD computing is apparent in the elements that the computer computes with.
In traditional computing the elements are Booleans, numbers, and memory pointers.
In HD computing they are multicomponent hypervectors, or tuples, where neither individual component nor a subset thereof has a specific meaning: a component of a hypervector and the entire hypervector represent the same thing.
Furthermore, the hypervectors are ultra-wide: the number of components is in the thousands and they are independent and identically distributed (i.i.d.).

We will demonstrate the idea with a simple example from language~\cite{Aditya,Rahimi2016}.
The task is to identify the language of a sentence from its three-letter sequences called \emph{trigrams}.
We compare the trigram profile of a test sentence to the trigram profiles of 21 languages and chose the language with the most similar profile.
A profile is essentially a histogram of trigram frequencies in the text
in question.

The standard algorithm for computing the profile---the \emph{baseline}---scans through the text and counts the trigrams.
The Latin alphabet of 26 letters and the space give rise to $27^3$ =
19,683 possible trigrams, and so we can accumulate the trigram counts
into a 19,683-dimensional vector and compare such vectors to find the
language with the most similar profile.
This is straightforward and simple with trigrams but it gets complicated with higher-order $n$-grams when the number of possible $n$-grams grows into the millions (the number of possible pentagrams is $27^5$ = 14,348,907).
The standard algorithm generalizes poorly.

The HD algorithm starts by choosing a set of 27 letter hypervectors at
random.
They serve as \emph{seed hypervectors,} and the same seeds are used with all training and test data.
We have used 10,000-dimensional hypervectors of equally probable $0$s and $1$s as seeds (aka binary spatter coding~\cite{BSC96}).
From these we make trigram hypervectors by \emph{rotating} the first letter hypervector twice, the second letter hypervector once, and use the third letter hypervector as is, and then by \emph{multiplying} the three hypervectors component by component.
Such trigram hypervectors resemble the seed hypervectors in that they are
10,000-$D$ with equally probable $1$s and $0$s, and they are random
relative to each other.
A text's profile is then the sum of all the trigrams in the text: for each occurrence of a trigram in the text, we \emph{add} its hypervector into the profile hypervector.
The profile of a test sentence is then compared to the language profiles and the most similar one is returned as the system's answer, as above.
In contrast to the standard algorithm, the HD algorithm generalizes readily to any $n$-gram size: the hypervectors remain 10,000-$D$.

\subsection{Arithmetic Operations on Hypervectors}
\label{sec:arith_ops}
HD computing is based on the properties of hypervectors and operations on them.
We will review them with reference to $D$-bit hypervectors, where $D$=10,000 for example~\cite{Kanerva98SDM}.
There are $2^D$ such hypervectors, also called \emph{points}, and they correspond to the corners of a $D$-dimensional unit cube.
The number of places at which two binary hypervectors differ is called the \emph{Hamming distance} and it provides a measure of \emph{similarity} between hypervectors.
A peculiar property of HD spaces is that most points are relatively far from any given point.
Hence two $D$-bit hypervectors chosen at random are dissimilar with near certainty: when referenced from the center of the cube they are nearly \emph{orthogonal} to each other.

To combine hypervectors, HD computing uses three operations~\cite{HD09}: \emph{addition} (which can be weighted), \emph{multiplication}, and \emph{permutation} (more generally, multiplication by a matrix).
``Addition'' and ``multiplication'' are meant in the abstract algebra sense where the sum of binary vectors $[A + B + \ldots]$ is defined as the
componentwise majority function with ties broken at random, the product is defined as the componentwise \verb|XOR| (addition modulo 2 denoted by \XOR), and permutation ($\rho$) shuffles the components.
All these operations produce a $D$-bit hypervector, and we collectively call them as Multiply-Add-Permute (MAP) operations~\cite{MAP98}.

The usefulness of HD computing comes from the nature of the MAP operations. Specifically, addition produces a hypervector that is \emph{similar} to the argument hypervectors---the inputs---whereas
multiplication and random permutation produce a \emph{dissimilar}
hypervector; multiplication and permutation are \emph{invertible}, addition is approximately invertible; multiplication \emph{distributes} over addition; permutation distributes over both multiplication and
addition; multiplication and permutation \emph{preserve similarity},
meaning that two similar hypervectors are mapped to equally similar hypervectors elsewhere in the space.

Operations on hypervectors can produce results that are approximate or ``noisy'' and need to be identified with the exact hypervectors.
For that, we maintain a list of known (noise-free) seed hypervectors in a so-called \emph{item memory} or \emph{clean-up memory.}
When presented with a noisy hypervector, the item memory outputs the most similar stored hypervector.
High dimensionality is crucial to make that work reliably~\cite{Kanerva98SDM}.
With 10,000-bit hypervectors, $1/3$ of the bits can be flipped at random and the resulting hypervector can still be identified with the original stored one.

These operations make it possible to encode/decode and manipulate sets, sequences, and lists---in essence, any data structure.
Such packing and unpacking operations are then viewed as mappings between points of the space suggesting a mechanism for analogy, with the analogy mapping being computed from examples.
This enables to implement analogical reasoning to answer non-trivial queries, e.g., ``What’s the Dollar of Mexico?''~\cite{Dollar-Mexico}.

\begin{figure}[t]
  \centering
  \includegraphics[width=0.6\columnwidth]{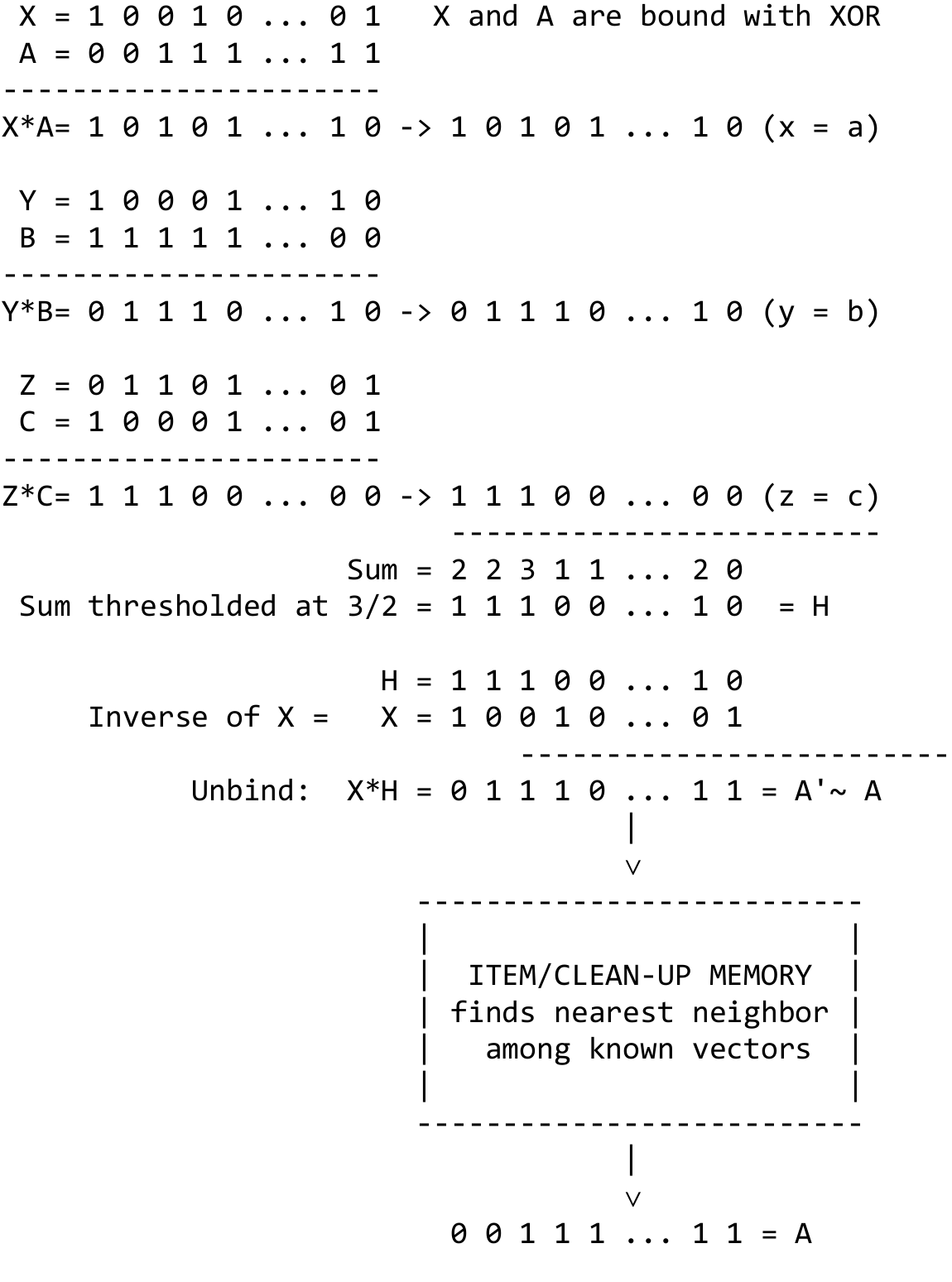}
  \caption{An example of encoding and decoding of a data structure using HD computing.}
  \label{fig:OPs}
\end{figure}
Figure~\ref{fig:OPs} shows how a data record consisting of variables $x, y, z$ with values $a, b, c$ can be encoded into a hypervector $H$ and the value of $x$ can be extracted from it. We start with randomly chosen seed hypervectors $X, Y, Z, A, B, C$ for the variable and the values and store them in the item memory. We then encode the record by \emph{binding} the variables to their values with multiplication and by adding together the bound pairs:
$$
   H = [(X \oplus A) + (Y \oplus B) + (Z \oplus C)]
$$
To find the value of $x$ in $H$ we multiply it with the inverse of
$X$, which for \verb|XOR| is $X$ itself: $A' = X \oplus H$. The
resulting hypervector $A'$ is given to the item memory which returns $A$ as the most-similar stored hypervector. An analysis of this example would
show how the properties of the operations, as listed above, come to
play. A thing to note about the operations is that addition and multiplication approximate an algebraic structure called a field, to
which permutation gives further expressive power.

HD computing has been described above in terms of binary hypervectors. However, the key properties are shared by hypervectors of many kinds, all of which can serve as the computational infrastructure. They include Holographic Reduced Representations (HRR)~\cite{PlateBook}, Frequency-domain Holographic Reduced Representations (FHRR)~\cite{PlateBook}, Binary Spatter Codes (BSC)~\cite{BSC96}, Multiply-Add-Permute (MAP)
coding~\cite{MAP98}, Binary Sparse Distributed Codes
(BSDC)~\cite{Rachkovskij2001}, Matrix Binding of Additive Terms (MBAT)~\cite{Gallant}, and Geometric Analogue of Holographic Reduced Representations (GAHRR)~\cite{GAHRR}. Different representational schemes using high-dimensional vectors and operations on them are generally referred to as Vector Symbolic Architectures (VSA)~\cite{VSA03} and the ultrahigh dimensionality is referred to as
\emph{hyperdimensional}~\cite{HD09}.

\subsection{General and Scalable Model of Computing}
\label{sec:universal}
\begin{figure}[t]
\begin{center}
	\subfigure[Text analytics]{\epsfig{file=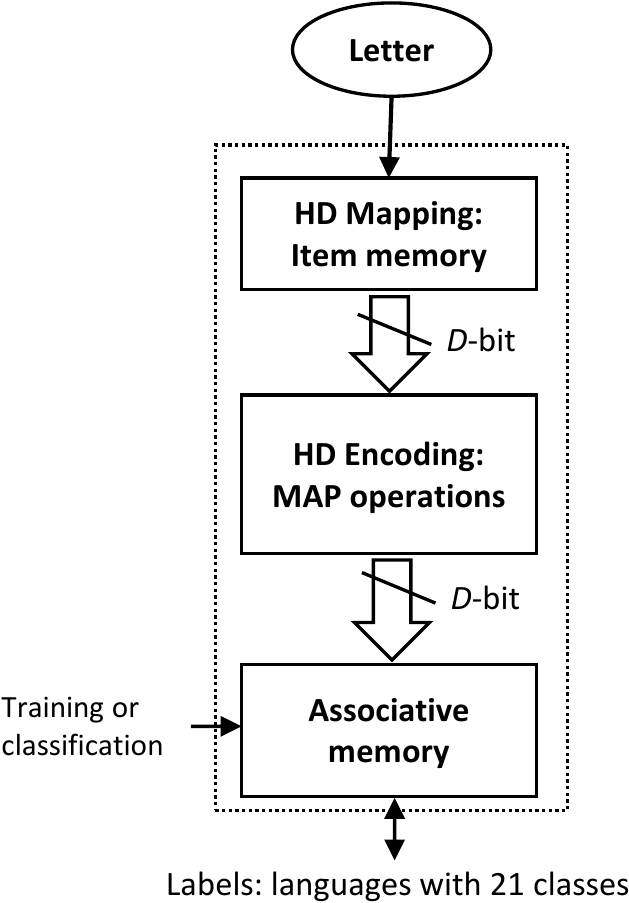, width=0.27\columnwidth}\label{fig:Language-arch}}
	\subfigure[EMG signals]{\epsfig{file=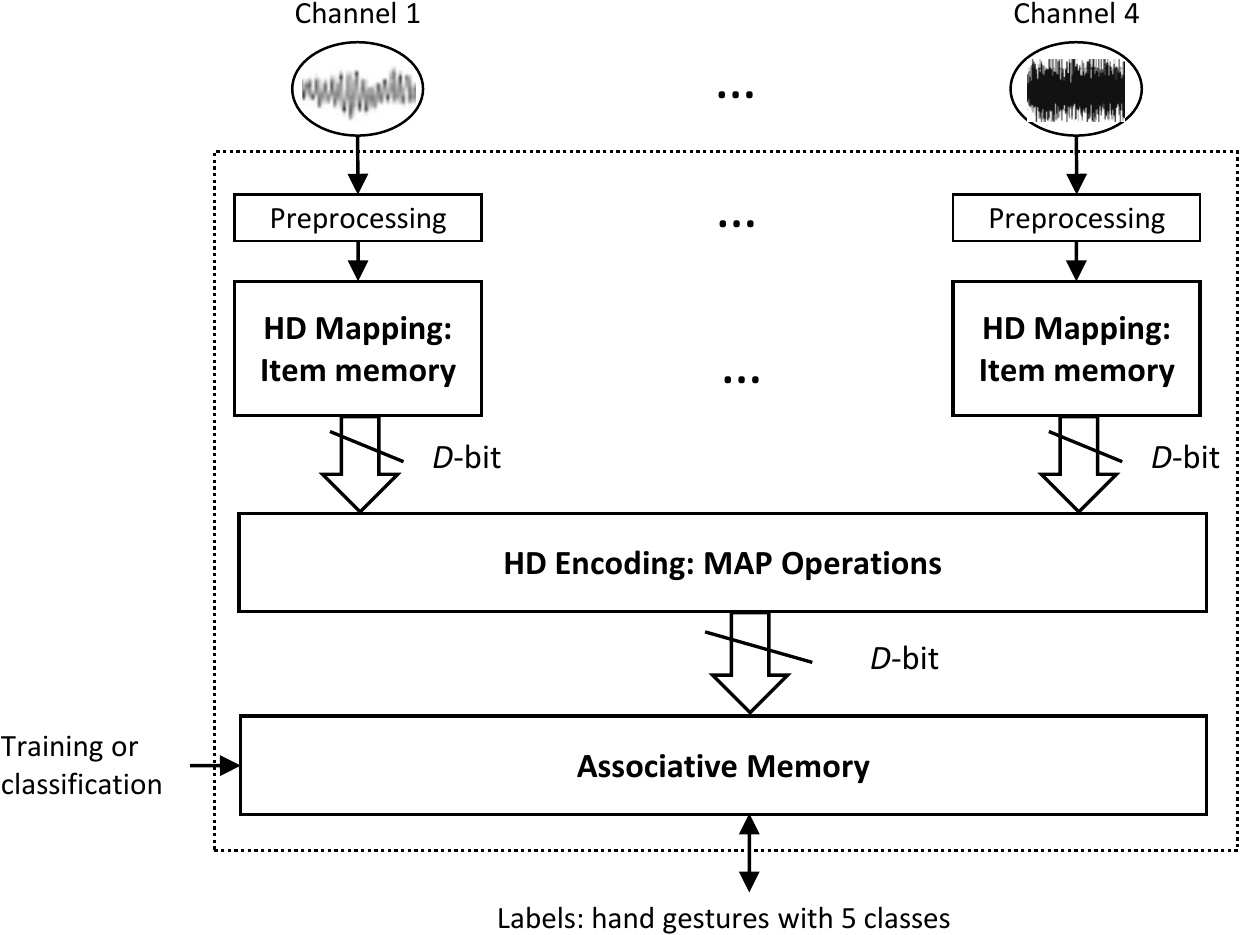, width=0.52\columnwidth}\label{fig:EMG-arch}}
\caption{General and scalable HD computing for various cognitive tasks: (a) European languages recognition; (b) EMG-based hand gesture recognition.}
\label{fig:universal}
\end{center}
\end{figure}
\begin{table}[t]
\centering
\caption{List of applications based on HD computing with different number of inputs and classes. The last two columns compare the classification accuracy of HD versus the baseline in that application domain.}
\label{table:apps}
\begin{tabular}{|l|l|l|l|l|}
\hline
Applications                     & Inputs (\#) & Classes (\#) & HD (\%) & Baseline (\%) \\ \hline
Language recognition~\cite{Aditya,Rahimi2016}             & 1           & 21           & 96.7\%           & 97.9\%                 \\ \hline
Text categorization~\cite{Rasti2016}              & 1           & 8            & 94.2\%           & 86.4\%                 \\ \hline
Speech recognition~\cite{VoiceHD_ICRC17}               & 1           & 26           & 95.3\%           & 93.6\%                 \\ \hline
EMG gesture recognition~\cite{EMG-HD}          & 4           & 5            & 97.8\%           & 89.7\%                 \\ \hline
Flexible EMG gesture recognition~\cite{ISCAS18} & 64          & 5            & 96.6\%           & 88.9\%                 \\ \hline
EEG brain-machine interface~\cite{BICT17,MONET17}      & 64          & 2            & 74.5\%           & 69.5\%                 \\ \hline
ECoG  seizure detection~\cite{BioCAS18}          & 100         & 2            & 95.4\%           & 94.3\%                 \\ \hline
\end{tabular}
\end{table}
HD computing is a complete computational paradigm that is easily
applied to learning problems.
Its main difference from other paradigms is that it can operate with data represented as approximate patterns, allowing it to scale to large learning applications.
HD computing has been used commercially since 2008 for making semantic vectors for words---semantic vectors have the property that words with similar meaning are represented by similar vectors.
The Random Indexing (RI)~\cite{Randomindexing} algorithm for making semantic hypervectors was developed as an alternative to Latent Semantic Analysis (LSA)~\cite{LSA97}, which relies on compute-heavy Singular Value Decomposition (SVD).
The original experiment used 37,000 ``documents'' on 7 topics to compute 8,000-dimensional semantic hypervectors of equal quality for 54,000 words.
SDV-based LSA requires memory in proportion to the product: `size of vocabulary' $\times$
`number of documents'.
By contrast, RI requires memory in proportion to the size of the vocabulary, and the statistics of documents/contexts is learned through simple vector addition~\cite{Randomindexing}.
Thus, the complexity of the method grows linearly with the size of the training corpus and scales easily to millions of documents.

Multiplication and permutation make it possible to encode causal relations and grammar into these hypervectors, thereby capturing more and more of the meaning in language~\cite{Aditya, RPRSKJ2015}.
We have used HD computing successfully to identify the language of test sentences, as described at the beginning of this section~\cite{Aditya,Rahimi2016} (also with sparse hypervectors~\cite{LangSpar_DT17}), to categorize News articles~\cite{Rasti2016}, and to classify DNA~\cite{HDNA}; other applications to text include common substrings search~\cite{BICA14} and recognition of permuted words~\cite{BICA16CT}.
HD computing has been also used in speech recognition~\cite{Rasanen2015con,VoiceHD_ICRC17}.
All these applications have a single input stream (Figure~\ref{fig:Language-arch}), while HD computing provides a natural fit for applications with multiple sensory inputs, e.g., biosignal processing (Figure~\ref{fig:EMG-arch}).
For instance, we have adapted the architecture for text analytics to the classification of hand gestures, when analog electromyography (EMG) signals are recorded simultaneously by four sensors~\cite{EMG-HD,PULP-HD,TNNLS18} or even a larger flexible electrode array~\cite{ISCAS18}.
The template architecture is shown in Figure~\ref{fig:universal}.
The architecture was further extended to operate on electroencephalography (EEG)~\cite{BICT17,MONET17}, and electrocorticography (ECoG)~\cite{BioCAS18} with up to 100 electrodes.

Notably, the learning and classification tasks are performed on the same hardware construct by integrating the following three main steps: 1) mapping to the HD space, 2) encoding with the MAP operations, and 3) associative memory (see Figure~\ref{fig:universal}).
The only difference is that during training, the associative memory updates the learned patterns with new hypervectors, while during classification it computes distances between a query hypervector and learned patterns.
Hence, it is possible to build a general-purpose computational engine based on these operations to cover a variety of tasks with similar success rates.
We show later in Section~\ref{sec:expdemo}, how such computational engine can be efficiently realized by using emerging nanotechnologies.
In addition, since the same hardware is used for learning and classification, the architecture is ideal for incremental or online learning.
\subsection{Robustness of Computations}
\label{sec:robustness}
HD computing is extremely robust.
Its tolerance for low-precision and faulty components is achieved by brain-inspired properties of hypervectors: (pseudo)randomness, high-dimensionality, and fully distributed holographic representations.
Symbols represented with hypervectors begin with i.i.d. components and when combined with the MAP operations, the resulting hypervectors also appear as
identically distributed random hypervectors, and the independence of the individual components is mostly preserved. This means that a failure in a component of a hypervectors is not ``contagious.'' At the same time, failures in a subset of components are compensated for by the holographic nature of the data representation, i.e., the error-free components can still provide a useful representation that is similar enough to the original hypervector. This inherent robustness eliminates the need for asymmetric error protection in memory units. This makes HD data representation suited for operation at low
signal-to-noise ratios (SNR).
\subsection{Memory-centric with Embarrassingly Parallel Operations}
\label{sec:memory-centric}
At its very core, HD computing is about manipulating and comparing large patterns within the memory itself.
The MAP operations allow a high degree of parallelism by needing to communicate with only a local component or its immediate neighbors.
Other operations such as the distance computation can be performed in a distributed fashion~\cite{HAM_HPCA17}.
This is a fundamental difference from traditional computational architectures, where data has to be transported to the processing unit and back, creating the infamous memory wall.
In HD processing, logic is tightly integrated with the memory and all computations are fully distributed.
This translates into substantial energy savings, as global interconnects are accessed at a relatively low frequency.
\section{Case Study: Language Recognition}
\label{sec:lang_example}
\begin{figure*}[t]
\centerline{
\epsfig{file=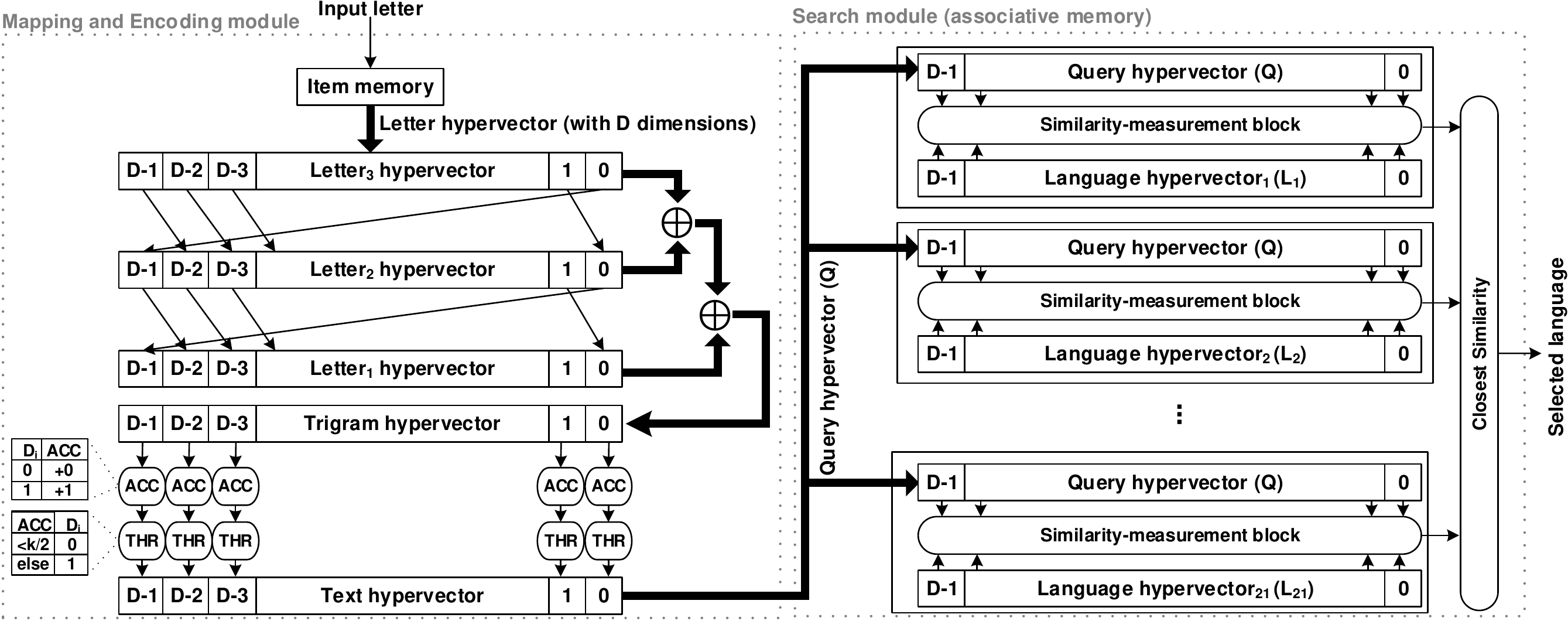, width=\linewidth}}
\caption{A 2D architecture of HD computing for language recognition task (see also Figure~\ref{fig:Language-arch}): mapping and encoding module, and search module.}
\label{fig:2D-architecture}
\end{figure*}
As a concrete application of HD computing, let us look at an
implementation of the language recognition algorithm discussed in Section~\ref{sec:HD}.
The HD algorithm generates trigram profiles as hypervectors and compares them for similarity.
As shown in Figure~\ref{fig:2D-architecture}, the design is based on a memory-centric architecture where logic is tightly integrated with the memory and all computations are fully distributed.
The HD architecture has two main parts: mapping and encoding modules, and similarity search module (associative memory).
The mapping and encoding module embeds an input text, composed of a stream of letters, to a hypervector in the HD space.
Then this hypervector is broadcast to the similarity-search module for comparison with a set of \emph{learned} language hypervectors.
Finally, the search module returns the language that has the closest match based on Hamming distance similarity.
\subsection{Mapping and Encoding Module}
\label{sec:encoding}
This module accepts the text as a stream of letters and
computes its representation as a hypervector.
The module has an item memory that holds a random hypervector (the ``letter'' hypervector) for each of the 26 letters and the space.
The item memory is implemented as a lookup table that remains constant.
In the dense binary coding~\cite{BSC96}, a letter hypervector has an approximately equal number of randomly placed $1$s and $0$s, hence the 27 hypervectors are approximately orthogonal to each other.
As another alternative, mapping to binary hypervectors can be realized by \emph{rematerialization}~\cite{HD_Man_module} e.g., by using a cellular automaton exhibiting exhibiting chaotic behaviour~\cite{CA+HD}.

The module computes a hypervector for each block of 3
consecutive letters as the text streams in. It consists of 3 stages
in FIFO style, each of which stores a letter hypervector. A trigram
hypervector is created by successively permuting the letter vectors
based on their order and binding them together, which creates a unique representation for each unique sequence of three letters.  For
example, the trigram ``abc'' is represented by the hypervector
$\rho(\rho(A) \oplus B) \oplus C=\rho(\rho(A)) \oplus \rho(B) \oplus
C$.  Use of permutation and binding distinguishes the sequence ``abc'' from ``acb'', since a permuted hypervector is uncorrelated with all the other hypervectors.

The random permutation operation $\rho$ is fixed and is implemented as a rotation to right by 1 position as shown in Figure~\ref{fig:2D-architecture}. For instance, given the trigram
``abc'', the $A$ hypervector is rotated twice ($\rho(\rho(A))$), the
$B$ hypervector is rotated once ($\rho(B)$), and there is no rotation
for the $C$ hypervector.  Once ``c'' is reached, its corresponding $C$ hypervector is fetched from the item memory and is written directly to the first stage of the encoder (i.e., Letter$_3$ hypervector in Figure~\ref{fig:2D-architecture}). The two previous letters are rotated as they pass through the encoder and turn into $\rho\rho(A)$ and $\rho(B)$. Componentwise bindings (i.e., multiplication) are then applied between these three hypervectors to compute the trigram hypervector, i.e., $\rho\rho(A) \oplus \rho(B) \oplus C$. Since the trigram hypervector is binary, the binding between two hypervectors is implemented with $D$ \verb|XOR| gates.

The hypervector for the input text is computed by adding together the
hypervectors for all the trigrams in the text and by applying a
threshold to retain them in the binary space. An input text of length $k+2$ generates $k$ trigram vectors. We implement the componentwise addition with a set of $D$ accumulators (ACC in Figure~\ref{fig:2D-architecture}), one for each
dimension of the hypervector, and count the number of $1$s in that component location. This componentwise accumulation produces a $D$-dimensional hypervector of integers.  To compute the corresponding binary hypervector, the encoding module applies a threshold of $k/2$ (implementing the majority function ($k$, $k/2$)) to every accumulator
value, where $k$ is the number of trigrams accumulated from the input. Left side of Figure~\ref{fig:2D-architecture} shows such a dedicated accumulation and thresholding for every hypervector component. The output of the module is the binary text hypervector.

The mapping and encoding module is used for both training and testing. During
training when the language of the input text in known, we refer to the text hypervector as a \emph{language} hypervector. Such language
hypervectors are stored in the search module as learned patterns. When the language of a text is unknown, as it is during testing, we call the text hypervector
a \emph{query} hypervector.  The query hypervector is sent to the
similarity search module to identify its source language.

\subsection{Similarity Search Module}
\label{sec:search}
The search module stores a set of language hypervectors that are precomputed by the mapping and encoding module.
These language hypervectors are formed in exactly the same way as described above, by making the text hypervectors from samples of a known language.
Therefore, during the training phase, we feed texts of a known language to the mapping and encoding module and save the resulting text hypervector as a language hypervector in the search module.
We consider 21 European languages and at the end of training have 21
language hypervectors, each stored in its own row of the search
module.

The language of an unknown text is determined by comparing its query
hypervector to all the language hypervectors. This comparison is done in a distributed fashion using an associative memory, and with the
Hamming distance as the similarity function.
Hamming distance counts the number of components at which two binary
hypervectors disagree.
The module uses a set of $D$ \verb|XOR| gates to identify mismatches between the two hypervectors.
In this digital implementation, the similarity-measurement block compares only one component each clock cycle. Hence, it takes $O(D)$ cycles to compute the Hamming distance between the two hypervectors~\cite{Rahimi2016}.
This block is replicated 21 times (the number of languages in our application) within the search module as shown in Figure~\ref{fig:2D-architecture}. The query hypervector is broadcast across the search module, hence all the
similarity-measurement blocks compute their distance concurrently.
Finally, a combinational comparison block selects the minimum Hamming
distance and returns its associated language as the language that the
unknown text has been written in.

\section{Emerging Technologies for HD Computing}
\label{sec:emergingTech}
Several emerging nanotechnologies such as carbon nanotube field-effect transistors, resistive RAM, and monolithic 3D integration have been demonstrated to be particularly effective for implementation of HD computing as well as other computing paradigms.
One or more of these technologies has been used in demonstrating HD computing operations~\cite{IEDM_2016} as well as in full system demonstrations~\cite{ISSCC2018,JSSC2018}. Here, we introduce these technologies.

\subsection{Carbon Nanotube Field-Effect Transistors}
Carbon nanotube field-effect transistors (CNFETs) are an emerging transistor technology which promises an order of magnitude improvement in energy-delay-product (a metric of energy efficiency) for digital circuits~\cite{Chang12}.
CNFETs use multiple carbon nanotubes (CNTs), which are cylindrical structures of carbon atoms 1-2 nm in diameter, that act as channels. CNTs enable highly energy-efficient digital logic circuits due to their high carrier mobility and excellent electrostatic control in CNFETs \cite{Appenzeller08}.
High-performance complementary logic has been demonstrated using CNFETs with an I$_{\textup{on}}$/I$_{\textup{off}}$ ratio (i.e., the ratio of drive current to the off-state leakage current) of about $10^6$~\cite{Shulaker14, Yang17}. They can be built at scaled gate lengths (5 nm)~\cite{Qiu271} and without hysteresis~\cite{Park17}.
CNFETs have even been fabricated as negative capacitance FETs with sub-55 mV/decade subthreshold swing at room temperature \cite{Srimani18}. CNFETs can be fabricated at low temperature ($\le$ 250$^{\circ}$C), which is key to enabling monolithic 3D integration (discussed in later in this chapter).

When designing circuits with CNFETs, imperfections inherent in CNTs, such as mis-positioned CNTs (that can lead to stray conducting paths resulting in incorrect functionality) and metallic CNTs (i.e., CNTs with little or no bandgap), can be overcome using the imperfection-immune paradigm. The imperfection-immune paradigm uses a combination of fabrication and design techniques~\cite{Zhang12, Shulaker15, Hills15} to enable wafer-scale fabrication and VLSI-compatible design of CNFET circuits. This paradigm has enabled experimental demonstrations such as the first CNT computer, the first 3D nanosystem consisting of over 2 million CNFETs on a single die, and the first full-system demonstration of an HD computing nanosystem~\cite{Shulaker13, Shulaker17, ISSCC2018}.

In addition to process variations that exist in silicon transistors (e.g., variations in threshold voltage, channel length, and oxide thickness), CNFETs are subject to CNT-specific variations such as CNT count variations (i.e., variations in the number of CNTs in a CNFET). These variations cause drive current variations, which can manifest as delay variations in digital circuits. These variations can be suppressed using optimized process and circuit design.
However, these inherent variations can be utilized in HD computing to generate the seed hypervectors, as demonstrated in~\cite{ISSCC2018} by capitalizing on the variations in CNT count and threshold voltage.
This means that variability and randomness essentially become the sources for computation.

\subsection{Resistive RAM}	
Resistive RAM (RRAM) is an emerging memory technology that promises high capacity, non-volatile data storage (10-year retention), and can be fabricated at low temperature ($\le$ 250$^{\circ}$C)~\cite{Metal-Oxide-RRAM, RRAM-nature}.
RRAM is fabricated as a metal oxide switching layer (insulator) sandwiched between two metallic electrodes and can be realized using various metal-insulator-metal material combinations.

Three main operations are typically performed on an RRAM cell: set, reset, and read. The set operation transforms the cell from high-resistance state (HRS) to low-resistance state (LRS) by applying a positive voltage (i.e., set voltage) across the top and bottom electrodes~\cite{Metal-Oxide-RRAM}. A transistor is typically used to limit the current for the set operation (called the compliance current). This creates or lengthens a filament of oxygen vacancies from the bottom electrode to the top electrode. As the length of the conductive filament increases, the resistance of the RRAM decreases~\cite{Metal-Oxide-RRAM}. In most cases, a higher set voltage (called forming voltage) is applied to form the filament (once) after fabrication. However, forming-less RRAM cells have also been demonstrated~\cite{Metal-Oxide-RRAM, Fang11}. A reset operation transforms the cell from LRS to HRS by applying a negative voltage (i.e., reset voltage) across the top and bottom electrodes, rupturing the filaments between the electrodes. RRAM cells with $\le$2 V set/reset voltage ($\approx$10 ns pulse duration) and 10-100 HRS/LRS resistance ratio have been demonstrated~\cite{Metal-Oxide-RRAM,Kim11}. RRAM is also subject to variations in its resistance, stemming from the stochastic size and shape of the conductive filament after a set or reset operation. These variations can also be exploited to generate seed hypervectors discussed in Section~\ref{sec:3D-VRRAM}. A read operation detects the state of the cell (e.g., HRS or LRS) by sensing the current after applying a small voltage across the two electrodes. This voltage is small enough to not change the resistance of the cell. Although RRAM has limited write (i.e., set/reset) endurance ($10^{12}$ cycles at the cell level~\cite{Kim11} and $10^5$-$10^7$ cycles at the array level~\cite{Grossi16, Chen17}), HD computing is shown to be robust against such endurance-related errors (see Section~\ref{sec:3D-VRRAM}).

Many cell structures (e.g., 1 transistor-1 RRAM cell, 1 transistor-n RRAM cell, 1 selector-1 RRAM cell~\cite{Metal-Oxide-RRAM}) may be used for RRAM, with each structure exhibiting a trade of between cell density (i.e., the number of cells that can be placed in a given area) and the controlability (of the resistance during set or reset operations) or the ability to detect the state of the cell reliably. For example, the 1 transistor-1 RRAM (1T-1R) cell configuration can be effectively used to prevent current overshoot during the set operation and provide exceptional selectivity between cells during the read operation but has limited cell density due to each cell using a transistor (typically larger than the RRAM cell itself)~\cite{Metal-Oxide-RRAM}. Array-level implementations using the 1T-1R RRAM structure have been demonstrated up to 16 Gbits of capacity~\cite{Fackenthal14}. Moreover, RRAM can be vertically built (3D VRRAM) in a bit-cost scalable manner to improve the cell density~\cite{ReRAM3D-inmemory}. HD computing demonstrations have used both the 1T-1R~\cite{ISSCC2018} configuration as well as 3D VRRAM~\cite{IEDM_2016}.

A single RRAM cell can store a single bit or multiple bits~\cite{Sheu11}. To demonstrate multi-bit storage in RRAM cells, one or a combination of set or reset parameters are adjusted to change the resistance of the cell to an intermediate value (between LRS and HRS): compliance current in the set operation, reset voltage, and set or reset pulse duration. These parameters can also be adjusted to gradually increase the RRAM cell resistance (i.e., increasing the resistance incrementally).
This gradual increase in RRAM cell resistance has been demonstrated for a variety of switching layers (i.e., the material in which the filament forms, such as HfO$_X$~\cite{Cabout13}) by using short pulses during the reset operation. This behavior, called \emph{gradual reset}, can be employed to realize addition operations in hardware~\cite{ISSCC2018}.

The RRAM has been demonstrated as digital storage, and as incrementers using gradual reset (i.e., the ability to increment the RRAM resistance in a fine-grained manner), and for performing the bitwise operations necessary for HD computing~\cite{ISSCC2018, IEDM_2016}.

\subsection{Monolithic 3D Integration}
Monolithic 3D integration is a process whereby tiers of circuits (i.e., a layer of logic, memory, or sensors) are fabricated on top of each other on the same substrate. Monolithic 3D integration uses inter-layer vias (ILVs),  standard vias used to connect adjacent metal layers in the interconnect stack of today's silicon CMOS technologies, to connect between tiers of circuits. This is in contrast to chip stacking using through-silicon vias (TSVs) with typical pitches of around 10 $\mu$m~\cite{Leduc08}. ILVs can have the same pitch as metal interconnects (100 nm at 28 nm technology node~\cite{Liang11}), enabling significantly denser vertical connectivity compared to TSVs~\cite{Batude11}---a key to tight integration between logic and memory.

Monolithic 3D integration requires low temperature fabrication for upper tiers of circuits ($\le$400$^{\circ}$C) as higher temperatures can damage existing circuits (transistors and interconnects) on the bottom tiers. While this is difficult for traditional silicon CMOS technologies (e.g., high temperature requirements for dopant activation $\ge$1,000$^{\circ}$C), it is naturally enabled by CNFETs and RRAM due to their low temperature fabrication~\cite{Metal-Oxide-RRAM, Patil09}. In recent demonstrations of HD computing, all CNFETs and RRAM were fabricated with a maximum temperature of 200$^{\circ}$C. Monolithic 3D integration of CNFETs, RRAM, and silicon transistors has been shown~\cite{Shulaker14}, demonstrating compatibility with silicon CMOS.
HD computing has been shown to provide up to 35$\times$ energy-execution time product benefits when implemented using such monolithically integrated CNFETs and RRAM compared to the standard silicon CMOS approach~\cite{JSSC2018}.

\section{Experimental Demonstrations for HD Computing}
\label{sec:expdemo}
In this section, we describe several experimental hardware demonstrations for HD computing, including in-memory MAP kernels using 3D VRRAM~\cite{IEDM_2016} and an end-to-end HD system with CNFETs and RRAM using monolithic 3D integration~\cite{ISSCC2018,JSSC2018}. Electrical characterization results and system simulations will be discussed to provide insights into how emerging device technologies can be utilized towards efficient implementations of HD computing systems.

\begin{figure}[t!]
  \centering
  \includegraphics[width=0.95\columnwidth]{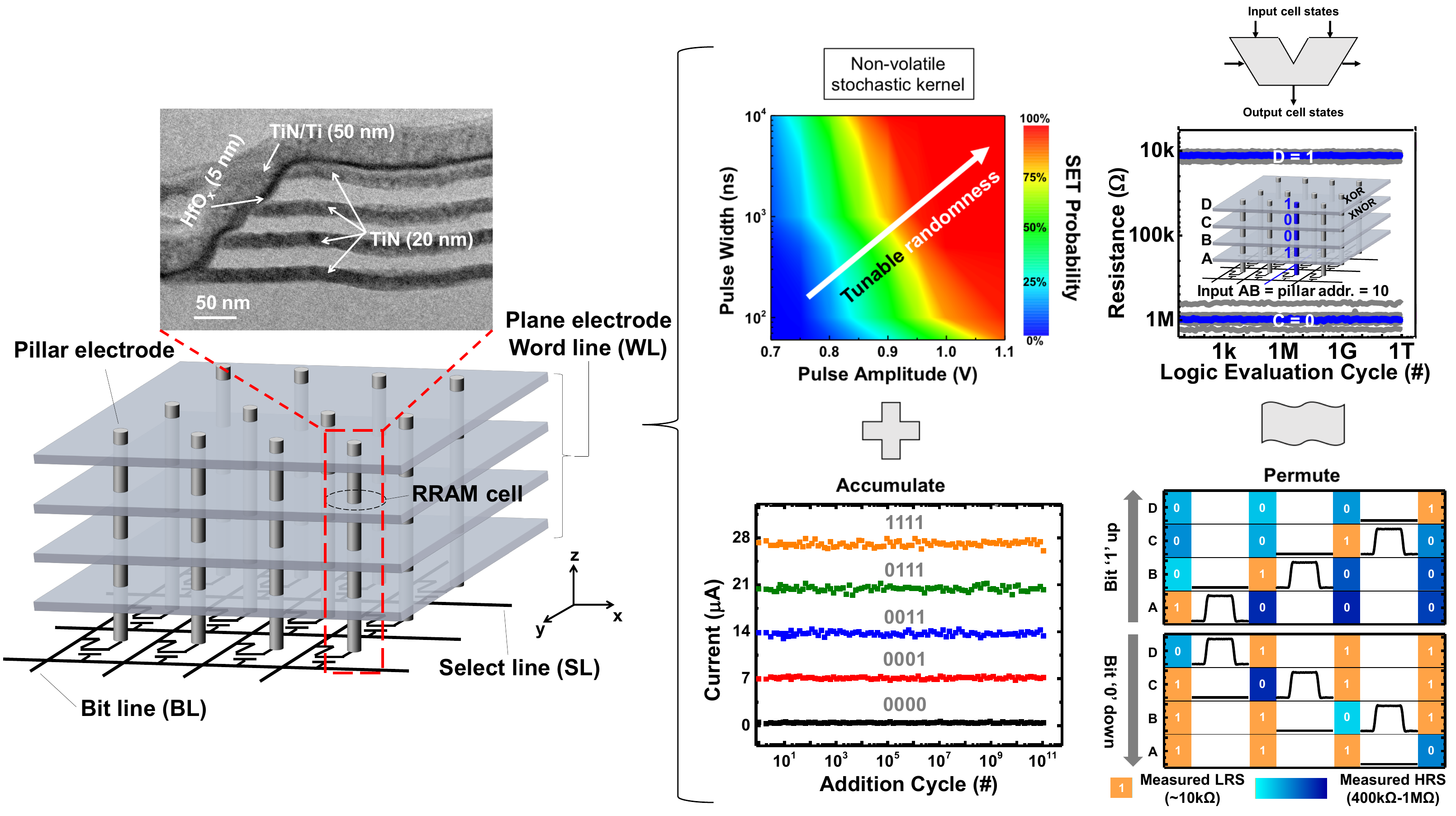}
  \caption{In-memory MAP kernels utilizing the unique 3D structure and non-volatility of 3D VRRAM. Electrical measurement data are collected from 4-layer 3D VRRAM devices.}
  \label{fig:VRRAM_MAP}
\end{figure}

\subsection{3D VRRAM Demonstration: In-Memory MAP Kernels}
\label{sec:3D-VRRAM}
\begin{figure}[t!]
  \centering
  \includegraphics[width=0.5\columnwidth]{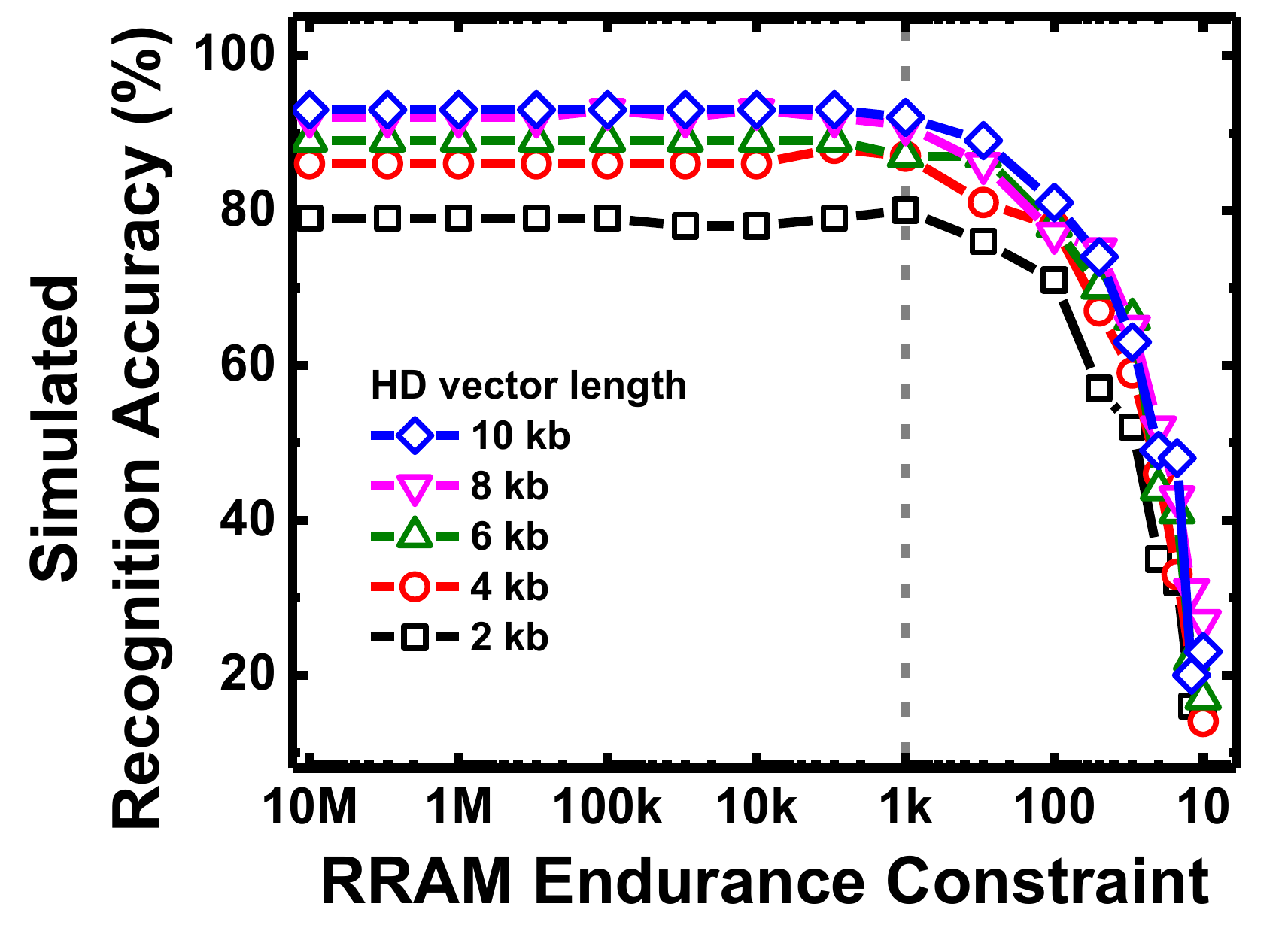}
  \caption{Evaluation of the impact of RRAM endurance constraints (switching-induced stuck-at errors).}
  \label{fig:VRRAM_endurance}
\end{figure}
One novel approach to realize memory-intensive MAP operations is to directly construct a set of native MAP kernels within a dense memory array without moving data around. 3D VRRAM, with multiple layers vertically stacked in a NAND-like fashion, has been found naturally suitable for the MAP kernel implementation (see Figure~\ref{fig:VRRAM_MAP}). Binary bits in HD representation are initialized and stored in RRAM cells in a non-volatile fashion, utilizing the native stochastic switching behaviors of RRAM. With a 50\% probability of switching from high-resistance state (HRS, representing `0') to low-resistance state (LRS, representing `1'), seed hypervectors can be generated. The central idea of performing MAP logic functions within 3D VRRAM is to utilize the 3D vertical connectivity and the non-volatility during modified write and read operations on 3D VRRAM. Figure~\ref{fig:VRRAM_MAP} shows an example of a 4-layer 3D VRRAM (i.e., a vertical 1T-4R structure) that is used to demonstrate the MAP operations~\cite{IEDM_2016}.

To yield arbitrary Boolean logic functions (including the bit-wise \verb|XOR|), voltage division between the select transistor and multiple RRAM cells (in this example, 4 RRAM cells) along a vertical pillar in 3D VRRAM is utilized during pulsing operations.
The voltage drop on certain RRAM cells is a function of resistance states of the rest of RRAM cells, thus creating a logic mapping from inputs to outputs. Detailed pulse operation schemes can be found in~\cite{li2017resistive}.
In the following, we describe how this method can be exploited to implement the MAP operations for HD computing (see also Figure~\ref{fig:VRRAM_MAP}):
\begin{itemize}
\item For multiplication, after one-time logic mapping, the 3D non-volatile memories serve as an \verb|XOR| look-up table, where bit-wise \verb|XOR| inputs are used to decode and read out the \verb|XOR| outputs stored in 3D VRRAM cells, eliminating the need of re-writing any RRAM cell. Up to $10^{12}$ \verb|XOR| evaluation cycles are performed by electrical measurement without disturb errors in readout results (Figure~\ref{fig:VRRAM_MAP}).

\item The addition kernel employs the analog-domain current summing property of 3D VRRAM. Along a vertical pillar which is `shared' by multiple layers of RRAM cells, current summing can be enabled by performing `read' operations in parallel on multiple layers. Proof-of-concept 4-bit vectors are written into multiple vertical pillars in 4-layer 3D VRRAM, and accumulated current during readout measurements correspond to correct outputs, with up to $10^{11}$ addition cycles measured (Figure~\ref{fig:VRRAM_MAP}).

\item Implementing permutation within 3D VRRAM does not require separate read-out and write-back processes. Similar to realizing Boolean functions, vertical connectivity (and voltage division) can be exploited to perform bit-wise data transfer among RRAM cells. Using pairs of VDD/GND pulses along vertical pillars, data transfer is implemented to perform permutation for hypervectors in a sequential fashion.
\end{itemize}

Since using non-volatile memory cells for both HD data representation and MAP operations involves write operations, endurance constraint (total number of write cycles before a hard error is produced) is also evaluated by conducting simulations on the language recognition task. During the training and inference on the task dataset, endurance failures (stuck at `1' or `0') may occur on RRAM cells. Under different levels of endurance characteristics at the device level, simulations show that certain degree of robustness can be retained (Figure~\ref{fig:VRRAM_endurance}), owing to the robust HD representation. Future work is needed on more realistic endurance modeling to enable device-system co-optimization being aware of RRAM endurance trade-offs.

\begin{figure}[t]
  \centering
  \includegraphics[width=0.75\columnwidth]{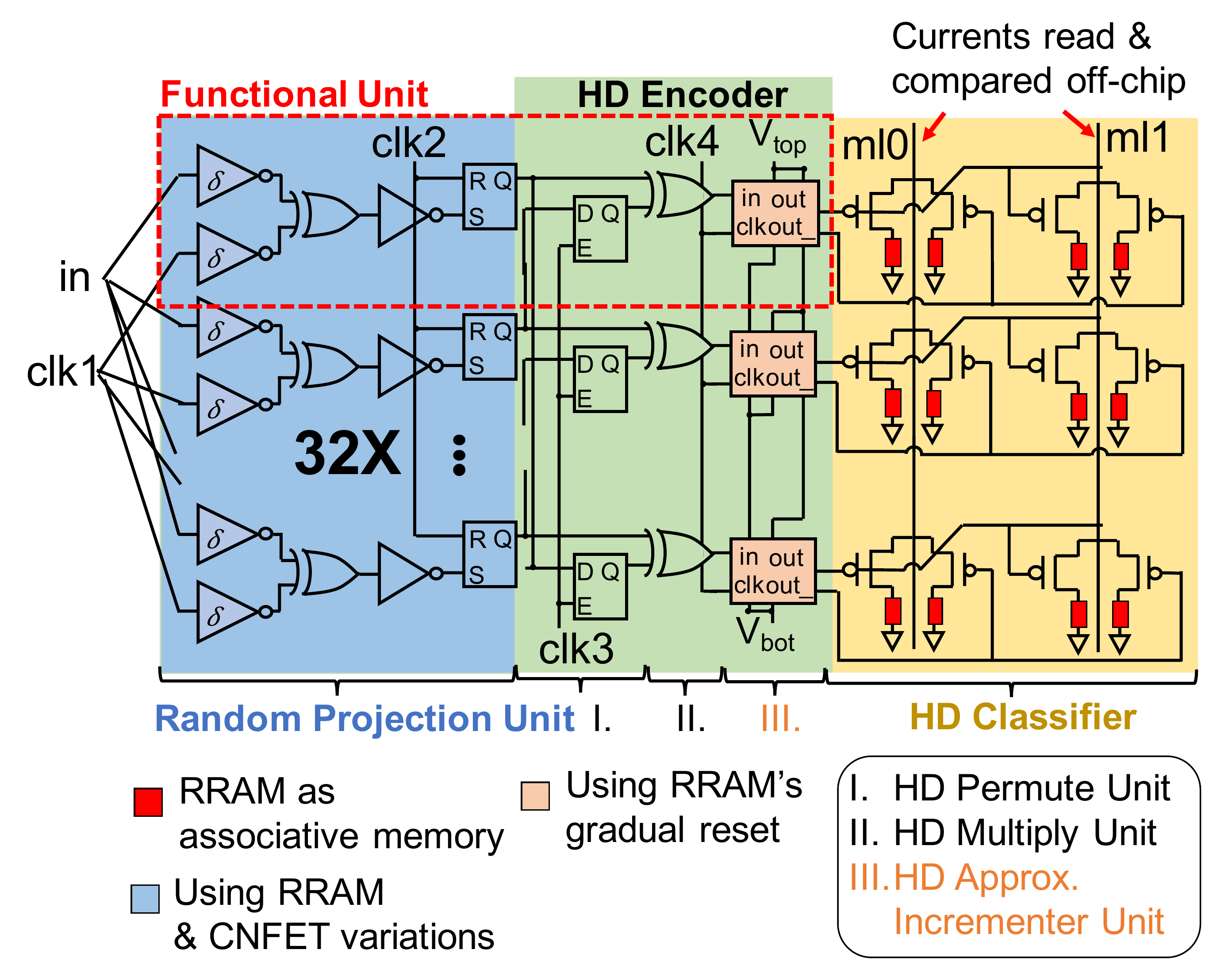}
  \caption{Schematic of Monolithic 3D HD system using CNFETs and RRAM.}
  \label{fig:HD_nanosystem}
\end{figure}
\subsection{System Demonstration using Monolithic 3D Integrated CNFETs and RRAM}
HD computing can be realized in hardware using monolithic 3D integration of CNFETs and RRAM~\cite{ISSCC2018,JSSC2018} and has demonstrated pairwise classification of 21 languages with measured mean accuracy of up to 98\% on $>$20,000 sentences. Unique properties of RRAM and CNFETs can be exploited to create area- and energy-efficient monolithic 3D circuit blocks that combine CNFETs with fine-grained access to RRAM memories (Figure~\ref{fig:HD_nanosystem}):
\begin{enumerate}
    \item Circuits that embrace inherent variations in RRAM and CNFETs, with estimated 3$\times$ lower dynamic energy (vs. silicon CMOS implementations at the same technology node) stemming from both the circuit topology and the use of energy-efficient CNFETs.
    \item Approximate incrementer circuits using gradual RRAM reset operation, which can use 30$\times$ fewer transistors vs. full-digital incrementer implementations.
    \item Ternary content addressable memory (TCAM) cells built using pairs of CNFETs and RRAM, that use 19$\times$ lower energy (simulated vs. SRAM-based TCAM cells) due to reduced leakage of non-volatile RRAM.
\end{enumerate}
When such a system is implemented at smaller technology nodes (e.g., 28nm node), it can simultaneously achieve lower energy and faster execution time compared to conventional silicon CMOS approaches (e.g., 7.6$\times$ lower energy and 4.6$\times$ faster execution time).

\begin{figure}[t]
  \centering
  \includegraphics[width=0.5\columnwidth]{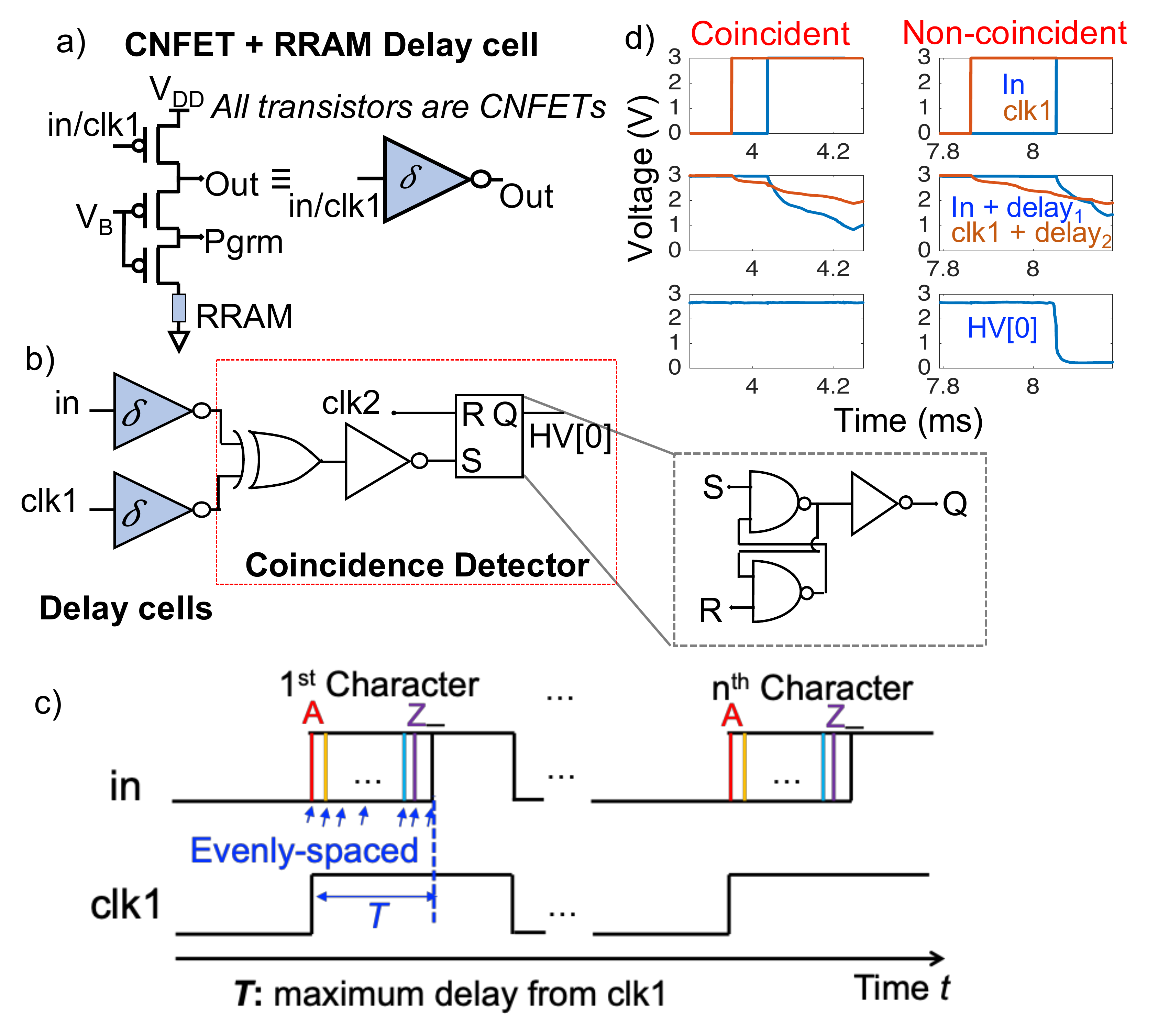}
  \caption{Delay cells exploiting inherent variations of CNFETs and RRAM}
  \label{fig:Delay_cells}
\end{figure}
\begin{figure}[t]
  \centering
  \includegraphics[width=0.5\columnwidth]{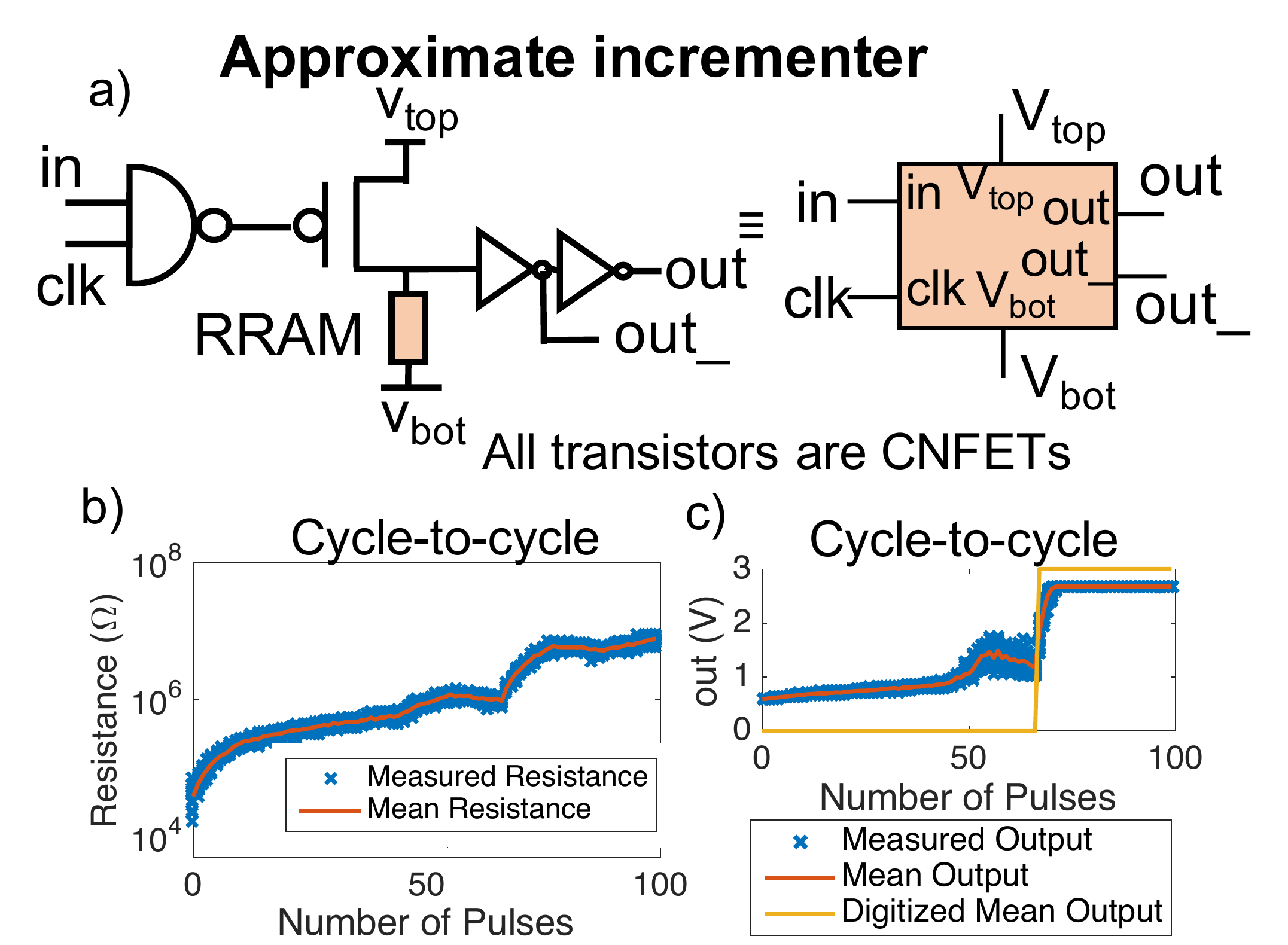}
  \caption{Approximate incrementer using the gradual reset property of RRAM. All transistors are CNFETs}
  \label{fig:incrementer}
\end{figure}

To realize an item memory, with randomly generated seeds, to map input letters to hypervectors, inherent variations in RRAM and CNFETs can be exploited (Figure~\ref{fig:Delay_cells}). Delay cells can be used to translate device-level variations such as drive current variations resulting from variations in carbon nanotube (CNT) count (i.e., the number of CNTs in a CNFET) or threshold voltage of CNFETs, and resistance variations of RRAM to delay variations.
To generate hypervectors, each possible input (26 letters of the alphabet and the space character) is mapped to a delay from a reference clock edge (time-encoded). To calculate each bit of the hypervector, random delays are added to both the input signal and the reference clock using delay cells. If the resulting signals are coincident (the falling edges are close enough to set a SR latch) the output is `1’. Before training the system, to initialize delay cells, the RRAM resistance is first reset to a high-resistance state (HRS) and then set to a low-resistance state (LRS).

To realize circuits to perform addition in hardware, an approximate incrementer with thresholding which leverages the multiple values of RRAM resistance that can be programmed by performing a gradual reset is used (Figure~\ref{fig:incrementer}). A digital buffer is used to transform (threshold) the sum to a binary hypervector. Each such approximate incrementer uses 8 transistors and a single RRAM cell. In contrast, a digital 7-bit incrementer may use 240 transistors. Thus, when $D$ (e.g. 10,000) accumulators are needed, the savings can be significant.

The search module is implemented using 2T2R (2-CNFET transistor, 2-RRAM) ternary content-addressable memory (TCAM) cells to form an associative memory (Figure~\ref{fig:HD_nanosystem}). During training, the matchline (i.e. ml0 or ml1) corresponding to the language (e.g., ml0 for English and ml1 for Spanish) is set to a high voltage to write to the RRAM cells (e.g., 3V), writing the query hypervector into the RRAM cells connected to the matchline. During inference, the matchlines (i.e. ml0 and ml1) are set to a low voltage (e.g. 0.5 V), and the current on each matchline is read as an output. When the query hypervector bit is equal to the value stored in a TCAM cell (match), the current is high. Otherwise (mismatch), the current is low. Cell currents are summed on each matchline. The line with the most current corresponds to the output class.


\end{document}